\documentclass[aps,prd,twocolumn,floatfix,showpacs,graphicx,print]{revtex4-1}
\usepackage{graphicx}
\usepackage{dcolumn}
\usepackage{bm}
\usepackage{epsfig}
\usepackage{lipsum}
\usepackage{natbib}
\usepackage[fleqn]{amsmath}
\usepackage{mathrsfs}
\usepackage{amssymb}
\usepackage{times}
\usepackage{psfrag}
\usepackage[english]{babel}
\usepackage{multirow}
\usepackage{color}
\begin{document}

\title{On the Mediating Field in a Conformally Transformed Einstein Equation}

\author{Dor Gabay}
\email[]{dorgabay@mail.tau.ac.il}
\affiliation{Department of Physical Electronics, Tel-Aviv University, Tel-Aviv 69978, Israel.}

\author{Sijo K. Joseph}
\email[]{sijo@wellgreen.co.in}
\affiliation{WellGreen Technologies Pvt. Ltd., Thodupuzha, Kerala, India.}

\begin{abstract}
A unique constraint is defined within the framework of scalar-tensor theories, whereby the conformal factor is fixed to the fluctuation associated to the effective mass of the Hamilton-Jacobi equation for a Klein-Gordon field. 
The effective mass is extended to its exponential form to remove any ghost (energy) states. 
The constraint's Lagrange multiplier $\lambda$, referred to as the mediating field, is shown to act as a mediator between the scalar and tensor degrees of freedom. 
In its linear form, Heisenberg's uncertainty principle appears as a natural artifact of the mediating field. 
In its exponential form, the mediating field is shown to be bound, nonsingular, and of increasing significance for smaller masses. 
Furthermore, in acquiring the stress-energy tensors, the cosmological constant $\Lambda$ is formulated for a stationary solution of the particle density and mediating field. 
As a result, the mysterious variation in $\Lambda$ is properly evaluated from its cosmological value to that of an electron, from which a 77 order difference is obtained. 
In our final remarks, the mediating field $\lambda$ is suggested to be characteristic of the vacuum's energy density. 
\end{abstract}

\date{\today}


\maketitle 

\section{Introduction}
Quantum mechanics and the general theory of relativity are governed by different fundamental assumptions. 
While quantum theory canonically defines a set of field variables over a flat Minkowski space-time, general relativity displaces the gravitational forces onto a manifold structure using a space-time metric $g_{\mu\nu}$. 
Unlike the deterministic form of general relativity, quantum mechanics is inherently defined as a probabilistic theory with complex quantities and commutation relations corresponding to point-like entities. 
Therefore, determining a quantum theory that naturally subsumes the gravitational effects imposed by the general theory of relativity is nontrivial. 

For $U(1)$ symmetric fields, many of these difficulties can be circumvented by reformulating quantum mechanics into its geometric form~\cite{BohmI}. Interpreting classical trajectories using the de-Broglie-Bohm picture is one possible way for integrating classical and quantum theories. 
Classically, one must ensure the particle trajectories inherit the fluctuations arising from quantum mechanics. One possible way of incorporating such fluctuations is through a conformal factor. 
Nalikar and Padmanabhan~\cite{Narlikar1981,Tanu_PLA,*Tanu_Nature} studied a quantized version of conformal fluctuations associated to the spacetime geometry. 
Later on, Santamato~\cite{Santamato_Schrodinger} derived a modified Schr\"odinger equation by considering the scale-invariant Weyl theory. Thereafter, Sidharth~\cite{Sidharth_QM} derived the deBroglie wavelength using noncommutative geometry and the Bohr-Sommerfeld quantization rule, without  invoking Heisenberg's uncertainty principle. 
Following that, Shojai et. al \cite{Shojai_Article} defined a conformally transformed action characteristic of the relativistic deBroglie-Bohm equation. 

Many extension of the general theory of relativity have been considered in recent years~\cite{BeyondEinsteinGravity,WeylReview2017,Bergmann1968,fR_HamFormlation,Shojai2008,fofT1,fofT2,Bekenstein2011_TeVeS,BekensteinPRD_TeVeS,Moffat_STVG,MOG_Moffat}.
Weyl conformal gravity is especially interesting due to the fact that it is a higher derivative theory with certain advantages over Einstein's field equation~\cite{Mannheim2012,MannheimGalRot2013}.
In particular, it is a 4-derivative theory and it is renormalizable~\cite{Mannheim2012}. 

A major problem facing conformal gravity is that it is non-unitary, making it incompatible with our current understanding of quantum mechanics. 
As a resolution, Manheim and Bender~\cite{Bender1,Bender2,Manheim_PTSym} demonstrated that the Hermitian nature of quantum mechanics can be generalized to account for space-time reflection (i.e. PT-Symmetry) in assuring real eigenvalues and unitary behavior.
Assuming PT-Symmetry is a proper extension of Hermiticity, conformal gravity can be used to properly characterize field equations within an altered form of quantum mechanics. 
Another difficulty is that posed by Weinberg's no-go theorem~\cite{WeinbegCosmo89}, whereby stationary solutions for the scalar and tensor degrees of freedom cannot be simultaneously attained without fine-tuning. 

In this manuscript, Weinberg's no-go theorem is bypassed in adopting a unique constraint within a conformally transformed scalar-tensor action. 
The constraint utilized the Hamilton-Jacobi equation to fix the conformal factor to the fluctuation associated to the effective mass of a Klein-Gordon field. 
The Hamilton-Jacobi equation is further extended to its exponential form to ensure a positive-definite effective mass. 
The constraint's Lagrange multiplier $\lambda$ acts as a mediating field between the scalar and tensor degrees of freedom. 
The mediating field's equation of motion contains peculiar properties, including an intricate relationship to Heisenberg's uncertainty principle. 
In its exponential form, the stationary solution to the mediating field $\lambda$ is shown to be bound, nonsingular, and of decreasing effect from the quantum to classical regime (e.g. more intense for smaller masses). 
Using the derived stress-energy tensors, an analytical expression is defined for the cosmological constant $\Lambda$ and the alleged 60-120 order difference in $\Lambda$ is exemplified for an arbitrary mass $m$ ranging from that of the universe to a single electron. 
Finally, an interpretation is given for the mediating field $\lambda$ as being intrinsically related to the vacuum energy density.

\section{Theory}
The conformal theory of gravity~\cite{mannheim1994ctg,obrien2017ctg} is composed of a scalar field (i.e. conformal factor $\Omega$) attained by conformally transforming the metric tensor $g_{\mu\nu}\to\Omega^{2}g_{\mu\nu}$ within Einstein's field equation. 
The use of scalar fields within the general theory of relativity was first proposed by Brans-Dicke~\cite{BransDicke}, who used it in an attempt to dictate the spatio-temporal dependence of the gravitational constant $\kappa$.
The scalar field acts as a mechanism for incorporating Mach's principle into General Relativity, whereby the absolute rotation of an inertial frame-of-reference is governed by the distribution of matter. 
Weinberg~\cite{WeinbegCosmo89} later devised a no-go theorem which demonstrated that for a simple Lagrangian of the form 
\begin{eqnarray}
\mathcal{L}=e^{4\Omega}\sqrt{-g}\mathcal{L}_{0}[\sigma], \label{weinbergLag0}
\end{eqnarray}
and in the limit of a relatively uniform conformal factor $\Omega$, one cannot guarantee that stationary solutions of the scalar and tensor equations are guaranteed without fine-tuning the parameter(s) $\sigma$ of $\mathcal{L}_{0}[\sigma]$. It can therefore be stated that, in varying the Lagrangian with respect to the conformal factor and metric, there exists degrees of freedom in the latter which are not mediated by the former
\begin{eqnarray}
\frac{\partial\mathcal{L}}{\partial\Omega}=0\;\;\to\;\;\frac{\partial\mathcal{L}}{\partial g_{\mu\nu}}\ne0. \label{ScalarEq}
\end{eqnarray}
The aforementioned fine tuning problem arises within any conformally transformed Lagrangian $\mathcal{L}$. 
We demonstrate that Weinberg's no-go theorem can be bypassed using a Lagrange multiplier $\lambda$ with a particular constraint. 
The constraint acts as an effective \textit{mediator}, enforcing the metric and scalar field to conform in the aforementioned limit. 
An action, analogous to that of Eq.~(\ref{weinbergLag0}), takes the form
\begin{eqnarray}
\mathcal{L}&=&e^{4\Omega}\sqrt{-g}\mathcal{L}_{0}(\sigma)\nonumber \\
& & +\sqrt{-g}\lambda\Bigl(\Omega-\mathcal{Q}(\varphi,\varphi^{*},\nabla_{\mu}\varphi,\nabla_{\mu}\varphi^{*},...)\Bigr). \label{weinbergLag}
\end{eqnarray}
Here, the function $\mathcal{Q}$ must satisfy a certain set of conditions to be specified.
By varying the action with respect to $g_{\mu\nu}$ and $\Omega$, one arrives at tensor and scalar equations of motion, respectively
\begin{eqnarray}
T_{\mu\nu}=T_{\mu\nu}^{WB}-2\frac{\delta}{\delta g_{\mu\nu}}\Bigl(\lambda\, \mathcal{Q}(\varphi,\varphi^{*},\nabla_{\mu}\varphi,\nabla_{\mu}\varphi^{*},...)\Bigr) \label{einsteinEq}
\end{eqnarray}
\begin{eqnarray}
\frac{\partial\mathcal{L}}{\partial\Omega}=0 \implies (T^{\mu}_{\mu})^{WB} +\lambda=0. \label{scalarEq}
\end{eqnarray}
Here, $T_{\mu\nu}^{WB}$ is the stress-energy tensor resulting from varying the Lagrangian within Eq.~(\ref{weinbergLag0}). The constraint equation is given by,
\begin{eqnarray}
\frac{\partial\mathcal{L}}{\partial \lambda}=0 \implies \Omega=\mathcal{Q}(\varphi,\varphi^{*},\nabla_{\mu}\varphi,\nabla_{\mu}\varphi^{*},...). \label{CnstrEq0}
\end{eqnarray}
By further equating the trace of the tensor equation of motion to that of the conformal factor 
\begin{eqnarray}
\frac{\delta\mathcal{L}}{\delta\Omega}=\textup{Tr}\Biggl(\frac{\delta\mathcal{L}}{\delta g^{\mu\nu}}\Biggr), \label{Lvar}
\end{eqnarray}
one arrives at a dynamical equation of motion for $\lambda$
\begin{eqnarray}
\lambda=-2g^{\mu\nu}\frac{\delta}{\delta g_{\mu\nu}}\Bigl(\lambda \, \mathcal{Q}(\varphi,\varphi^{*},\nabla_{\mu}\varphi,\nabla_{\mu}\varphi^{*},...)\Bigr). \label{CnstrFieldEq1}
\end{eqnarray}
The equation of motion for the dynamical field variable $\lambda$ mediates the tensor degrees of freedom not present within the action of Eq.~(\ref{weinbergLag0}), resulting in a flat metric tensor in the limit of a uniform conformal factor 
\begin{eqnarray}
\frac{\partial\mathcal{L}}{\partial\Omega}=0\;\;\to\;\;\frac{\partial\mathcal{L}}{\partial g_{\mu\nu}}=0. \label{ScalarEq1}
\end{eqnarray}
The fine tuning problem can therefore be overcome by introducing an unexplored, nonzero field $\lambda$. For the Lagrange multiplier to act as a mediator, a certain set of conditions must be satisfied:
(i) The function $\mathcal{Q}$ must be composed of covariant derivatives to inherit a metric dependence. In the limit of a constant function $\mathcal{Q}=K$, the conformal factor $\Omega=K$ and the mediating field $\lambda=0$, rendering the constraint meaningless; (ii) The constraint within the action must be dimensionless; (iii) Given the conformal transformation is applied to the metric tensor $g_{\mu\nu}\to\Omega^2 g_{\mu\nu}$, the constraint should not be arbitrary, rather have some intrinsic dependence on the metric. 
In obeying the above conditions, the degrees of freedom not present in Eq.~(\ref{weinbergLag0}) can be further interpreted. 

To properly define the constraint, condition (iii) should be more rigorously articulated. In constraining the conformal factor, one would expect the function $\mathcal{Q}$ to be composed of the very entity which imposes the curvature within Einstein's equation. 
In dealing with a Klein-Gordon field, the associated relativistic quantum Hamilton-Jacobi equation takes the form
\begin{eqnarray}
\nabla_{\mu}S\nabla^{\mu}S=(1+Q)m^{2}=\Omega_{lin}^{2}m^{2}. \label{HJeqn_lin}
\end{eqnarray}
Here, $Q=\frac{\hbar^{2}}{m^{2}}\frac{\Box\sqrt{\rho}}{\sqrt{\rho}}$ is the relativistic quantum potential and is a result of adopting the Bohmian interpretation~\cite{BohmI,BohmII,Carroll2005,RCarroll2007}, whereby the scalar field $\varphi=\sqrt{\rho}\textup{e}^{iS/\hbar}$ is partitioned into its density $\sqrt{\rho}$ and effective phase $S$. Given $\nabla_{\mu}S\nabla^{\mu}S=E^{2}-p^{2}$ is simply the energy-momentum associated to the metric, constraining the conformal factor to its effective mass more rigorously satisfies condition (iii), while simultaneously conforming to conditions (i) and (ii). In assuring a positive-definite effective mass, one can further extend the linear conformal factor $\Omega_{lin}$ to its exponential form
\begin{eqnarray}
\begin{aligned}
\nabla_{\mu}S\nabla^{\mu}S&=(1+Q+\frac{1}{2}Q^{2}+...)m^{2} \\
&=\textup{e}^{Q}m^{2}=\Omega^{2}m^{2}. \label{HJeqn}
\end{aligned}
\end{eqnarray}
This removes any negative norm (ghost) states contained within the Klein-Gordon equation at the cost of no longer dwelling within its linear-order form.
The corresponding ghost states can easily be identified for spatio-temporal regions of the relativistic quantum potential $Q<-1$ within Eq.~(\ref{HJeqn_lin}), where the energy-momentum is bound to be negative. Unlike the linear order relation, the energy-momentum in Eq.~(\ref{HJeqn}) is guaranteed to be positive-definite. 

In defining $\mathcal{Q}$ and fully satisfying conditions (i)-(iii), we have fully bypassed Weinberg's 1989 no-go theorem. To proceed, we propose a conformally transformed action, composed of the Einstein-Hilbert Lagrangian $\mathcal{L}_{EH}$, a Klein-Gordon field, and the aforementioned constraint
\begin{widetext}
	\begin{eqnarray}
	A[g_{\mu\nu},\Omega,\varphi,\lambda]=\int\mathrm{d}^{4}x\sqrt{-g}\Biggl\{\frac{1}{2\kappa}\mathcal{L}_{EH}[g_{\mu\nu},\Omega^{2}]+\Omega^{2}\frac{\hbar^{2}}{m}\Bigl[\nabla_{\mu}\varphi^{*}\nabla^{\mu}\varphi-\frac{m^{2}}{\hbar^{2}}\Omega^{2}\varphi^{*}\varphi\Bigr]+\lambda\Bigl[\mathrm{ln}(\Omega^{2})-\mathcal{Q}(\varphi,\varphi^{*},\nabla_{\mu}\varphi,\nabla_{\mu}\varphi^{*},...)\Bigr]\Biggr\} \nonumber \\
	 \label{vacuaA}
	\end{eqnarray}
\end{widetext}

Here, the exponential conformal factor of Eq.~(\ref{HJeqn}) has been adopted, with the exponent replaced by the natural logarithm on the left-hand side of the constraint. The function $\mathcal{Q}$ is the scalar field equivalent of the relativistic quantum potential. One can imagine that expanding $\mathcal{Q}$ in terms of the scalar field results in a highly intangible expression. For this reason we choose to interpret the action of Eq.~(\ref{vacuaA}) in its Bohmian form. Such Bohmain actions have been thoroughly explored by Shojai and coworkers~\cite{Shojai2008,Shojai_Article}. Taking $c=1$, the Bohmian extension of Eq.~(\ref{vacuaA}) takes the form
\begin{eqnarray}
& & A[g_{\mu\nu},{\Omega}, S, \rho, \lambda]=\nonumber \\
& & \frac{1}{2\kappa}\int{d^4x\sqrt{-g}\left(R\Omega^2-6\nabla_{\mu}\Omega\nabla^{\mu}\Omega\right)}  \nonumber \\
& & +\int{d^4x\sqrt{-g} \left(\frac{1}{m}\rho\Omega^2 \nabla_{\mu}S \nabla^{\mu}S-m\rho\Omega^4\right)} \nonumber \\
& & +\int{d^4x\sqrt{-g}\lambda\left[\ln{\Omega^2}-\frac{\hbar^2}{m^2}\frac{\nabla_{\mu}\nabla^{\mu}\sqrt{\rho}}{\sqrt{\rho}}
\right]}. \label{Actioneq} 
\end{eqnarray}
The proposed action trivially reduces to the sum of Einstein-Hilbert action and the classical Hamilton-Jacobi equation in the $\hbar\to 0$ limit (i.e. $\Omega^2\to 1$). 
By minimizing the action with respect to $S, \rho, {\Omega}, g_{\mu\nu}$ and $\lambda$, we can further define the equations of motion for a relativistic matter field.
In first taking the variation of $A$ with respect to $\rho$, the equation of motion for the matter field is obtained  
\begin{eqnarray}
(\nabla_{\mu}S \nabla^{\mu}S- m^2\Omega^2)\Omega^2 \sqrt{\rho}+f(\lambda,\rho)=0. \label{EqMotion}
\end{eqnarray}
Here, $f(\lambda,\rho)=\frac{\hbar^2}{2m}[\Box({\frac{\lambda}{\sqrt{\rho}})}-\lambda\frac{\Box\sqrt{\rho}}{\rho}]$ is the contribution resulting from the constraint.
The equation of motion is fully defined in terms of the density $\rho$, phase (e.g. Hamilton's principal function) $S$, and Lagrange multiplier $\lambda$. 
In further taking the variation with respect to the phase $S$, one arrives at the corresponding continuity equation 
\begin{eqnarray}
\nabla_{\nu}(\rho\Omega^2\nabla^{\nu}S) =0. \label{ContinuityEq}
\end{eqnarray}
The quantum mechanical behavior of the matter field can therefore fully be described by two real fields $\rho$ and $S$, along with the yet undefined equation of motion for $\lambda$. 
In the limit of $\lambda\to\rho$, $f(\lambda,\rho)=0$ and one attains the usual Hamilton-Jacobi equation for a Klein-Gordon field. 
It is therefore apparent that without constraining the conformal factor, the Noether current corresponding to the continuity equation (Eq.~(\ref{ContinuityEq})) is not conserved and, as a result, the field could arbitrarily dissipate energy in a non-unitary manner. Imposing a constraint within such a scalar-tensor action can therefore be argued to be of important physical significance. 
The meaning of $\lambda$ is worthwhile to explore.
In this article, the Lagrangian multiplier $\lambda$ is interpreted as a mediator between the energy density of the scalar field and the stress-energy tensor of Einstein's equation. 
In varying the action with respect to $\lambda$, one obtains the exponential prescription of the conformal factor
\begin{eqnarray}
\Omega^2=\exp{\left(\frac{\hbar^2}{m^2}\frac{\nabla_{\mu}\nabla^{\mu}\sqrt{\rho}}{\sqrt{\rho}}\right)}. \label{CnstrEq}
\end{eqnarray}
This constraint equation is particularly interesting since we are identifying a purely geometrical quantity $\Omega^2$ with a quantum mechanical descriptor. 
As exemplified in Eq.~(\ref{CnstrFieldEq1}), the dynamical behavior of $\lambda$ can only be acquired in considering the metric dependence of the constraint. 
In varying the action with respect to $\Omega$, one obtains an equation of motion for the scalar curvature $R$ 
\begin{eqnarray}
R\Omega+6 \Box\Omega +\frac{2\kappa}{m}\rho \Omega (\nabla_{\mu}S \nabla^{\mu}S-2m^2\Omega^2)+\frac{2\kappa\lambda}{\Omega}=0. \nonumber \\
\label{TraceEq}
\end{eqnarray}
In further varying the action with respect to the metric tensor ${g}_{\mu\nu}$, the conformally transformed Einstein equation, along with its stress-energy tensors, are attained 
\begin{eqnarray}
\mathcal{G}_{\mu \nu}=T^{\bf{matter}}_{\mu\nu}(S,\rho)+T^{\bf{qm}}_{\mu\nu}(\Omega)+T^{\bf{med}}_{\mu\nu}(\lambda,\rho) 
\label{EinsteinEq}
\end{eqnarray}
\begin{eqnarray}
T^{\bf{matter}}_{\mu\nu}(S,\rho)&=& -\frac{2\kappa}{m}\rho \nabla_{\mu}S \nabla_{\nu}S 
 +\frac{\kappa}{m} \rho\,g_{\mu\nu}\nabla_{\sigma}S \nabla^{\sigma}S \nonumber \\
 & &-\kappa m \rho \Omega^2 g_{\mu\nu}. \label{tuv_matter}
\end{eqnarray}
\begin{eqnarray}
T^{\bf{qm}}_{\mu\nu}(\Omega) &=&\frac{(g_{\mu \nu}\Box{\Omega^2}- 
\nabla_{\mu}\nabla_{\nu}{\Omega^2})}{\Omega^2}+6\frac{\nabla_{\mu}\Omega 
\nabla_{\nu}\Omega}{\Omega^2} \nonumber \\
& & -3 g_{\mu \nu} \frac{\nabla_{\sigma}\Omega \nabla^{\sigma}\Omega}{\Omega^2}. \label{tuv_qm}
\end{eqnarray}
\begin{eqnarray}
T^{\bf{med}}_{\mu\nu}(\lambda,\rho)= 
-\frac{\kappa\hbar^2}{m^2\Omega^2}\Biggl[\nabla_{(\mu}&&\sqrt{\rho}\nabla_{\nu)}\Bigl(\frac{\lambda}{\sqrt{\rho}}\Bigr) \nonumber \\ 
&&-g_{\mu\nu}\nabla_{\sigma}\Bigl(\lambda\frac{\nabla^{\sigma}{\sqrt{\rho}}}{\sqrt{\rho}}\Bigr)\Biggr]. \label{TVacuum}
\end{eqnarray}
Here, the brackets in the index notation are symbolic of the symmetrized permutation of the respective indices. The resulting Einstein's equation is composed of stress-energy tensors related to the matter $T^{\bf{matter}}_{\mu\nu}(S,\rho)$, quantum $T^{\bf{qm}}_{\mu\nu}(\Omega)$ and mediating field $T^{\bf{med}}_{\mu\nu}(\lambda,\rho)$ contributions.
Clearly, the derivation of $T^{\bf{med}}_{\mu\nu}$ is nontrivial due to the presence of a second-order covariant derivative. Appendix~\ref{sec:app1} is therefore provided to thoroughly justify its result. Due to the presence of the Compton wavelength squared, the mediating field plays an ever increasing role in the limit of small mass. Therefore, the stress-energy tensor $T^{\bf{med}}_{\mu\nu}$ becomes critical in the quantum limit $m \to 0$ and tends to vanish in the classical limit $m \to \infty$. 

In understanding the impact of the constraint in Eq.~(\ref{Actioneq}) on Einstein's equation, it is important to interpret the mediating field $\lambda$. It is apparent that, for a real valued mediating field, $T^{\bf{med}}_{\mu\nu}(\lambda,\rho)$ results in negative curvature within the modified Einstein equation. 
Such behavior can be heuristically identified in studying the equation of motion for the scalar curvature $R$ in the limit of null particle velocity and mass
\begin{eqnarray}
R \approx -\frac{2\kappa}{\Omega^2}\lambda. \label{Rlimit}
\end{eqnarray}
A negative scalar curvature implies that, in the such a limit, Einstein's equation must correspondingly acquire negative energies. 
In general relativity, negative energies are well known to contribute to spacetime expansion~\cite{Guth_Inflation}. 
In addition to Brans-Dicke theories, $f(R)$ theories~\cite{fofr_agravity,EnrgyFRGravity,EnerfofrandBransDicke} similarly result in negative energies which contribute to an expanding physical metric. 
In quantum field theory, such negative local energy densities are attained by vacuum fluctuations, leading to sub-vacuum phenomena~\cite{ford2010negative}.
To study a more complete form of Eq.~(\ref{Rlimit}), a mathematical expression of the scalar curvature $R$ can be acquired by substituting Eq.~(\ref{EqMotion}) into Eq.~(\ref{TraceEq}),
\begin{equation}
R= +2\kappa\rho m {\Omega}^2 -6\frac{\Box\Omega}{\Omega} -\frac{2\kappa}{\Omega^2}\lambda
+\frac{\kappa \hbar^2 
\sqrt{\rho}}{m^2\Omega^2}{\Bigl(\Box{(\frac{\lambda}{\sqrt{\rho}})}-\lambda\frac{\Box{\sqrt{\rho}}}{\rho}\Bigr)}. \label{Req}
\end{equation}
As expected, the covariant derivative of the conformal factor $\Omega^2$ also negatively contributes to the scalar curvature $R$. 
Hence, as opposed to the positive-definite effective mass term in Eq.~(\ref{Req}), the expansion terms appearing in Eq.~(\ref{Rlimit}) are a manifestation of both the mediating field and conformal factor. 

The equation of motion for $\lambda$ can be obtained by substituting the scalar curvature of Eq.~(\ref{TraceEq}) into the contracted Einstein equation (Eq.~(\ref{EinsteinEq})), analogous to the procedure outlined in Eq.~(\ref{Lvar})
\begin{eqnarray}
2\nabla_{\alpha}\Bigl(\lambda\frac{\nabla^{\alpha}\sqrt{\rho}}{\sqrt{\rho}}\Bigr)
-\nabla_{\mu}\sqrt{\rho}\,\nabla^{\mu}\Bigl(\frac{\lambda}{\sqrt{\rho}}\Bigr)=\frac{m^{2}}{\hbar^{2}}\lambda.
\end{eqnarray}
The resulting partial differential equation can further be simplified into the following form
\begin{eqnarray}
\frac{m^{2}}{\hbar^{2}}\lambda(1-Q)=\nabla_{\sigma}\Bigl(\lambda\frac{\nabla^{\sigma}{\sqrt{\rho}}}{\sqrt{\rho}}\Bigr) \label{lambda_deq}.
\end{eqnarray}
In determining the solution of $\lambda$ perturbatively, expanding $\sqrt{\rho}$ and $\lambda$ for some small parameter trivially results in a null solution for all orders of $\lambda$, but such an approach is misleading. Eq.~(\ref{lambda_deq}) clearly contains nontrivial solutions which could arise from non-perturbative methods. To explore these potential solutions, we first separate the field variables on the right- and left-hand sides, while still preserving their covariant form
\begin{eqnarray}
 \frac{\nabla_{\mu}\sqrt{\rho}}{\sqrt{\rho}}\frac{\nabla^{\mu}\lambda}{\lambda}=\Bigl(\frac{m^2}{\hbar^2}(1-Q)
 -\frac{\Box\sqrt{\rho}}{\sqrt{\rho}}+\frac{\nabla_{\mu}\sqrt{\rho}\,\nabla^{\mu}\sqrt{\rho}}{\rho} \Bigr). \nonumber\\. 
\label{LambdaEq}
\end{eqnarray}
In order to explore the behavior of the mediating field, we first find a way to express $\lambda$ analytically in a fixed coordinate frame.
Adopting spherical coordinates, and assuming spherical symmetry (i.e. ignoring variations in $\theta$ or $\phi$), we arrive at the following equation of motion
\begin{eqnarray}
\partial_{t}\lambda=\Bigl(\frac{m^{2}}{\hbar^{2}}&&\frac{\sqrt{\rho}}{\partial_{t}\rho}(1-Q)-\frac{\Box{\sqrt{\rho}}}{\partial_{t}\sqrt{\rho}}+\frac{\partial_{\alpha}\sqrt{\rho}\,\partial^{\alpha}\sqrt{\rho}}{\sqrt{\rho}\partial_{t}\sqrt{\rho}}\Bigr) \lambda \nonumber \\
&&+(\frac{\partial_{r}\sqrt{\rho}}{\partial_{t}\sqrt{\rho}}) \partial_{r}\lambda.
\end{eqnarray}
Here, the covariant derivatives are simply partial derivatives with contracting indices still present and indicated by $\alpha$. The subscript $r$ denotes differentiation in the radial component only. 
In the case of a separable density $\rho(r,t)=\rho(r)\rho(t)$, it can be shown that the mediating field is of the form $\lambda =\lambda_{0} \exp{(-\int{dt\,\alpha_{1}(\rho(t))}+\int{dr \beta_{2}(\rho(r)})}$.
To further simplify matters, we study only stationary solutions of $\lambda$. In such a limit, the solution of $\lambda$ is analogous to that of a first-order partial differential equation
\begin{eqnarray}
\lambda(r)&=& \exp{\Bigl(-\int{dr\,\beta(r)+ {C}_{\beta} }\Bigr)} \label{LambdaRadEq} \\
\beta(r)&=&\frac{\sqrt{\rho}}{\partial_{r}\sqrt{\rho}}\Bigl(\frac{m^2}{\hbar^2}(1+Q_{r})
+\frac{\partial_{r}^{2}\sqrt{\rho}}{\sqrt{\rho}}-\frac{\partial_{r}\sqrt{\rho}\partial_{r}\sqrt{\rho}}{\rho}\Bigr).\nonumber \\ \label{Betaeq}
\end{eqnarray}
Here, $Q_{r}$ is the  radial contribution from the relativistic quantum potential $Q$ and $\lambda_{0}=\exp({C}_{\beta})$ is the resulting integration constant. The value of $\lambda_{0}$ can be determined by the boundary condition of choice, for which we do not prescribe.
Integrating over $\beta(r)$ within the exponential and computing ${C}_{\beta}$ results in a stationary solution for the mediating field. 
Once attained, interpreting the physical meaning of $\lambda$, including the aforementioned negative energies arising from a negative Ricci scalar, becomes a trivial task.

A formal justification of extending the linear-order constraint to that of an exponential one has not yet been provided. The advantage can be demonstrated by comparing the analytical solution to the stationary equations of motion for both approaches. We begin with the linear-order constraint. 
In instead adopting $\Omega_{lin}=(1+Q)$ with a Lagrange multiplier $\lambda_{lin}$, 
the mediating field's corresponding equation of motion is surprisingly similar to that of Eq.~(\ref{lambda_deq})
\begin{eqnarray}
\frac{m^{2}}{\hbar^{2}}\lambda_{lin}=\nabla_{\sigma}\Bigl(\lambda_{lin}\frac{\nabla^{\sigma}{\sqrt{\rho}}}{\sqrt{\rho}}\Bigr). \label{lambda_deq_linear}
\end{eqnarray}
The equation of motion for the linear-order mediating field takes the same form as that of Eq.~(\ref{LambdaRadEq})-(\ref{Betaeq}), with the exception of substituting $(1+Q)\to1$ in the first term of Eq.~(\ref{Betaeq}). 

Explicitly obtaining a solution for $\lambda$ and $\lambda_{lin}$, even using Eq.~(\ref{LambdaRadEq})-(\ref{Betaeq}), is not self-explanatory due to its density-dependence. The density is inherently a function of the wave equation defined by Eq.~(\ref{EqMotion}) and Eq.~(\ref{ContinuityEq}). The nontrivial form of these coupled equations of motion forces us to resort to ansatz procedures. Although ansatz procedures are heuristic, they provide a physical sense for the behavior of $\lambda$ per the form of $\sqrt{\rho}$. We choose to prescribe a stationary Gaussian distribution for the local density $\sqrt{\rho}$ in spherical coordinates
\begin{eqnarray}
\sqrt{\rho(r,s)} = {\bigl(\frac{1}{\pi\,s^2}\bigr)}^{3/4} \exp{\Biggl(\frac{-r^2}{2s^2}\Biggr)}. \label{dnstyExp}
\end{eqnarray}
Here, $N(s)={(\frac{1}{\pi\,s^2})}^{3/4}$ is the normalization constant and $s=\sigma+\sqrt{2}\ell_{p}$ is the spatial variation of the particle. 
In enforcing a stationary Gaussian solution of the density, we presume that there exists some finite external potential which could be added to the action of Eq.~(\ref{Actioneq}) at no cost of the analysis that has been articulated up to now. 
Although $\sigma$ can be chosen freely, a minimum length is assigned to comply with the Planck length $\ell_{p}$. 
As will be exemplified, the Gaussian's standard deviation $s$ cannot arbitrarily decrease to $\ell_{p}$, rather must stay above the minimum allowable standard deviation $\sqrt{2}\ell_{p}$. Here, the prescribed $\sqrt{2}$ is critical as it eliminates the singularity which will later appear in evaluating the cosmological constant from the derived stress-energy tensors.

Given the ansatz (Eq.~(\ref{dnstyExp})), the solution to the linear-order mediating field $\lambda_{lin}$ is
\begin{eqnarray}
\lambda_{lin}=\lambda_{0}\, r^{-(d-1)+ (s/{\ell_{c})^2}} \label{lambda_resultd}.
\end{eqnarray}
Here, $d$ is the number of space-time dimensions prescribed in fixing the coordinate system and $\ell_{c}=\hbar/m$ is the Compton wavelength of a particle of mass $m$. 
The linear-order mediating field $\lambda_{lin}$ clearly contains an asymptotic singularity at $s=\ell_{c}$, one which can be overcome when $s\geq \sqrt{(d-1)}\,\ell_{c}$.
Furthermore, in $1+1D$ spacetime ($d=2$), the free parameter of a normal distribution $s\propto\sqrt{2} \Delta{x}$ and the maximum uncertainty in momentum is $mc\propto\sqrt{2}\Delta{p}$ such that $\ell_{c}\approx\hbar/\sqrt{2}\Delta{p}$. In avoiding the singularity in Eq.~(\ref{lambda_resultd}), we can then arrive at a familiar relation
\begin{eqnarray}
\Delta x \Delta p \geq \,\hbar/2. \label{HUprinc}
\end{eqnarray}
This is just the uncertainty principle in the $1+1D$ scenario. A fundamental principle of quantum mechanics emerges from a stationary solution of the linear-order mediating field. 
Even when the condition of Eq.~(\ref{HUprinc}) is satisfied, $\lambda_{lin}$ increases polynomially to some order. The source-like contribution within the wave equation would then be infinite infinitely far away. This unphysical behavior suggests that $\lambda_{lin}$ would only suffice in the near-field regime of a local density. 
In the exponential extension of the linear-order theory, the singularity disappears and the polynomial increase of $\lambda_{lin}$ inherits an effective bound.
Using Eq.~(\ref{LambdaRadEq}) and Eq.~(\ref{Betaeq}) along with the predefined Gaussian density, the stationary solution of $\lambda$ takes the form
\begin{eqnarray}
\lambda=\lambda_{0}\exp{\Biggl(\frac{-r^2}{2s^2}\Biggr)} {r}^{(s/\ell_{c})^2} \label{eternal_eq}.
\end{eqnarray}
By ansatz, one can easily verify that this simple solution obeys Eq.~(\ref{lambda_deq}). As is apparent, the asymptotic behavior in Eq.~(\ref{lambda_resultd}) now decays exponentially, analogous to the prescribed Gaussian density. The mediating field $\lambda$ inherits a quasi-local behavior, such that it is nonzero in the domain neighboring the local regime of the density. Interestingly enough, in the classical limit $s\approx\sqrt{2}\ell_{p}$, the maximum $\lambda(r_{max})$ occurs exactly at the Schwarzschild radius $r_{max}=r_{s}$. Furthermore, Heisenberg's uncertainty principle is no longer inherent to the mediating field, rather some nontrivial extension to be studied in later works. 

We proceed to the study of the derived stress-energy tensors (Eq.~(\ref{tuv_matter})-(\ref{TVacuum})) within Einstein's field equation.
To better understand them, one must determine whether observables are properly reproduced in the quantum and cosmological regimes. 
One such observable is the cosmological constant $\Lambda$. 
The alleged 60-120 order difference in the transition of  $\Lambda$ from the classical to quantum regime is a long-lived problem yet to be solved~\cite{SeanCarrol_Cosmo,Zeldovich_Cosmo}. 
The massive difference is argued to result from vacuum fluctuations with cutoff energies at the Planck scale within quantum field theory. 
Typically, cosmological constants are determined under the assumption of the cosmological principle~\cite{barrow2005cosmoprinciple,weinberg1987cosmoprinciple}, whereby, in the large scale, the spatial distribution of the matter density is assumed to be relatively uniform and isotropic. Under the assumption of such a matter distribution, the Friedman-Lema$\hat{\textup{i}}$tre-Robertson-Walker (FLRW) metric~\cite{green2014flrw} is then adopted to characterize the temporal evolution of some scale factor $\textup{a}(t)$. 
In dealing with probabilistic, rather than bulk matter densities, we choose to approach the problem differently. Rather than articulating the distribution of matter, we approximate the probabilistic density of the bulk matter as a stationary normal distribution obeying the central limit theorem. In this framework, the probabilistic density is not the same as the matter distribution attained from the scale factor $\textup{a}(t)$. 
Dwelling in a time-independent frame-of-reference of the probabilistic density does not provide a comprehensive dynamical picture of the matter distribution, but, we hope, suffices in attaining the cosmological constant in some stationary limit of its comoving frame. 

The cosmological constant resulting from the proposed theory can be identified as the sum of all terms paired to $g_{\mu\nu}$ within the stress-energy tensors of Eq.~(\ref{tuv_matter})-(\ref{TVacuum})
\begin{eqnarray}
\Lambda_{v+q+g}&=& 3 \frac{\nabla_{\sigma}\Omega \nabla^{\sigma}\Omega}{\Omega^2} -\frac{\Box\Omega^2}{\Omega^2} 
- \frac{\kappa}{m} \rho\,\nabla_{\sigma}S \nabla^{\sigma}S \nonumber  \\
& &
+ \kappa m \rho \Omega^2 - \frac{\kappa}{\Omega^2}\lambda. \label{Lambda1}
\end{eqnarray}
Substituting the equation of motion (Eq.~(\ref{EqMotion})) into Eq.~(\ref{Lambda1}), one gets a quantity completely composed of the conformal factor, mediating field, and particle density
\begin{eqnarray}
\Lambda_{v+q+g}&=& 3 \frac{\nabla_{\sigma}\Omega \nabla^{\sigma}\Omega}{\Omega^2} -\frac{\Box\Omega^2}{\Omega^2} - \frac{\kappa 
\lambda (1-Q)}{\Omega^2}  \nonumber \\
& &
+\frac{\kappa\hbar^2\sqrt{\rho}}{2m^2\Omega^2}\Bigl[\Box({\frac{\lambda}{\sqrt{\rho}})}-\lambda\frac{\Box\sqrt{\rho}}{\rho}\Bigr]. 
\label{Cosmo}
\end{eqnarray}
Here, components in Eq.~(\ref{Cosmo}) composed of covariant derivatives of the conformal factor tend to be the dominant contribution of $\Lambda_{v+q+g}$. 
For particles of small mass, terms corresponding to the mediating field can potentially play a significant role in articulating expansion behavior. For $m$ on the order of the Planck mass $m_{p}=\sqrt{\hbar/G}$ or greater, the mediating field tends to play a less significant role then the conformal factor. 
Also, given that the mediating field is a quasi-local entity, with null contribution at the local regime of the density (i.e. the presence of nonlocal behavior solely in the neighborhood of a field's local domain), the local rate of expansion may be sufficiently governed by the first two terms of Eq.~(\ref{Cosmo}). 
By evaluating $\Lambda_{q+v+g}$ in Eq.~(\ref{Cosmo}) using the stationary normal distribution of Eq.~(\ref{dnstyExp}), conformal factor $\Omega^2$ (Eq.~(\ref{CnstrEq})), and analytically determined mediating field (Eq.~(\ref{eternal_eq})), one arrives at a complete expression for the cosmological constant
\begin{widetext}
\begin{eqnarray}
\Lambda_{v+q+g}= -6 \Bigl(\frac{\ell_{c}^2}{s^4}\Bigr)&&-\Bigl(\frac{\ell_{c}^2}{s^4}\Bigr)^2 r^2 -\kappa\,r^{s^2/\ell_{c}^2}\exp{\Biggl(\frac{-r^2}{2s^2}-\frac{\ell_{c}^2}{s^4} (r^2 -3 
s^2)\Biggr)}\Biggl(1+\frac{\ell_{c}^2}{s^4}(3s^2-r^2)\Biggr) \label{cosmo_secret} \\
&& +\frac{\hbar^2 \kappa}{2 m^2 \ell_{c}^4 s^4} r^{-2 +s^2/\ell_{c}^2}\exp{\Biggl(\frac{-r^2}{2s^2}-\frac{\ell_{c}^2}{s^4} (r^2 -3 
s^2)\Biggr)}\bigl[s^6 (\ell_{c}^2 + s^2)- \ell_{c}^4 (r^4 - 3 r^2 s^2) \bigr]. \nonumber
\end{eqnarray}
\end{widetext}
In the stationary limit, $\Lambda_{v+q+g}$ inherits a nontrivial spatial dependence which can be simplified by considering only its dominant contributions 
\begin{eqnarray}
\Lambda= 6 \Bigl(\frac{\ell_{c}^2}{s^4}\Bigr)+\Bigl(\frac{\ell_{c}^2}{s^4}\Bigr)^2 r^2. \label{CosmoSimpEq1}
\end{eqnarray}
The resulting expression contains a spatially quadratic radial component and a uniform contribution which dominates for $r\ll{s}^{2}/\ell_{c}$. 
Interestingly enough, in conformally transforming the action, the number $6$ results from the prescribed dimensionality of spacetime $(d=4) \implies (d-1)(d-2)=6$. 
The Gaussian variance obeys a minimum length requiring $s\geq\sqrt{2}\ell_{p}$ and is used to avoid $r^{-2}$ singularity in the third term of the Eq.~(\ref{cosmo_secret}). 
At $s=r=0$, this singularity $(\frac{1}{0})^0$ is corrected for a minimum length $s$, such that $(\frac{1}{0})^0\to{(\frac{1}{\ell_{p}})}^{\sqrt{2}\ell_{p}}$. 
In the cosmological scale, whereby $s \approx \sqrt{2}\ell_{p}\gg \ell_{c}$, Eq.~(\ref{CosmoSimpEq1}) can be expressed entirely in terms of the Schwarzschild radius $r_{s}=2Gm$,
\begin{eqnarray}
\Lambda \approx 6 \Bigl(\frac{1}{r_{s}^2}\Bigr)+\Bigl(\frac{1}{r_{s}^4}\Bigr) r^2. \label{CosmoAstroEq1}
\end{eqnarray}
For $r\ll{r_{s}}$, the static term dominates and can be shown to play a fundamental role in defining its astronomical value
\begin{eqnarray}
\Lambda_{\bf{astr}} \approx 6 \Bigl(\frac{1}{r_{s}^2}\Bigr). \label{CosmoAstroEq2}
\end{eqnarray}
Given an expression for the mass of the universe, the cosmological constant can then be analytically determined.
The Hoyle-Carvalho relation~\cite{Carvalho1995} provides a theoretical estimation for the mass of the universe using only microscopic quantities. 
D. Valev later pointed out that, using dimensionality arguments, the Hoyle-Carvalho relation can be approximated~\cite{DValev2014}
\begin{eqnarray}
M_{u} \propto \frac{c^3}{G\,H_{0}} \label{Massuni}.
\end{eqnarray}
Using Eq.~(\ref{CosmoAstroEq2}) and Eq.~(\ref{Massuni}), the cosmological constant can be written in terms of Hubble's constant $H_{0}$
\begin{eqnarray}
\Lambda_{\bf{astr}} \approx \frac{3}{2} \Bigl(\frac{H_{0}^2}{c^2}\Bigr). \label{CosmoAstroEq3}
\end{eqnarray}
Up to a factor of half, this is the standard form of the cosmological constant resulting from Friedmann's equation for a flat universe~\cite{Friedman1999} (i.e. when the gravitational mass density contribution is ignored). 
In starting with a pure quantum mechanical entity, we arrived at a standard result in cosmology differing only by a factor of half.
Given the recently measured Hubble constant ($H_{0} = 73.52 \pm 1.62\, km\, s^{-1}\, {Mpc}^{-1}$)~\cite{ExperiHubble2016,ExperiHubble2018}, the mass of the universe can be estimated $M_{u} \approx 1.6\times 10^{53} kg$.
By then inserting the mass of the universe into Eq.~(\ref{CosmoAstroEq2}), one arrives at a static contribution of $\Lambda_{\bf{astr}}=1.06\times{10}^{-52} m^{-2}$.  
Even with such a crude approximation, this value is unimaginably close to that measured in a recent experiment~\cite{PlanckMission}. 
The static term in Eq.~(\ref{CosmoAstroEq1}) therefore plays a fundamental role in characterizing the expansion of our universe. 
In further considering the quadratic polynomial term in Eq.~(\ref{CosmoAstroEq1}) at $r=r_{s}$ the rate of expansion naturally increases to $\Lambda=1.24\times{10}^{-52} m^{-2}$. 

In the small mass regime, the standard deviation within Eq.~(\ref{CosmoSimpEq1}) takes on a different form $s\approx \ell_{c}\gg \sqrt{2}\ell_{p}$, resulting in a quantum analogue for expansion
\begin{eqnarray}
\Lambda_{\bf{qm}}\approx 6 \Bigl(\frac{1}{\ell_{c}^2}\Bigr)+\Bigl(\frac{1}{\ell_{c}^2}\Bigr)^2 r^2. \label{LCosmoQM}
\end{eqnarray}
Similarly, ignoring the spatially dependent contribution, one arrives at a static value of the cosmological constant for an electron mass $\Lambda_{el}=4.02363\times{10}^{25}{m}^{-2}$. 
At the Compton wavelength of an electron $r=\ell_{c}$, the spatially quadratic radial term once again begins to dominate, resulting in $\Lambda_{el}=4.69424\times{10}^{25}{m}^{-2}$. This value may further change as a result of the increasing importance of the mediating field in the generalized expression of Eq.~(\ref{cosmo_secret}). 
The large discrepancy of the expansion in the small mass regime compared to its astronomical value is evident, equating to an unequivocable 77 orders.

\section{Summary \& Outlook}
In this paper, we have explored the implications of introducing a unique constraint into a conformally transformed scalar-tensor theory. 
The conformal factor was defined as the fluctuation associated to the effective mass of the Hamilton-Jacobi equation for a Klein-Gordon field. 
The constraint allowed us to bypass Weinberg's no-go theorem, and, in the process, introduce an equation of motion for the Lagrange multiplier $\lambda$. 
The Lagrange multiplier was suggested to act as a mediator between a Klein-Gordon field and Einstein's equation, and was hence nicknamed the mediating field. 
In adopting a normal distribution for the density as an ansatz, the behavior of the mediating field was studied in its stationary limit. 
It was found that, although the linear-order constraint resulted in a mediating field obeying Heisenberg's uncertainty principle, an exponential constraint was both bound and nonsingular. 
Furthermore, the cosmological properties of the stress-energy tensors were extensively studied. 
$\Lambda$ was shown to naturally vary by 77 orders from its cosmological to quantum scale and almost perfectly conform to the measured astronomical value. 

The sum of these results led us to speculate a peculiar correspondence between the mediating field $\lambda$ and the vacuum energy density. While, within quantum field theory, field variables are characterized by wave equations and quantized to interpret particle correlations in the presence of potentials, the alternative theory would extend these notions to a quantized wave-like field interacting with the proposed mediator. In this sense, the quantized field would endlessly seek to balance itself with its encompassing background and the background would constantly act as an interacting potential for the quantized field. The field theoretic consequences thereof will be explored in future works.

\appendix
\begin{widetext}
\section{Attaining $T_{\mu\nu}^{med}$} \label{sec:app1}
Varying second-order covariant derivatives is not common in the literature. This appendix is therefore provided for readers to properly understand how to account for the metric variation within the constraint $A_{cnstr}$ of Eq.~(\ref{Actioneq}). In first applying a chain rule for the metric variation, one attains
\begin{eqnarray}
\delta{A}_{cnstr}=\int{d}^{4}x\frac{\delta}{\delta{g^{\mu\nu}}}(\sqrt{-g})\lambda\left[\ln{\Omega^2}-\frac{\hbar^2}{m^2}\frac{\nabla_{\sigma}\nabla^{\sigma}\sqrt{\rho}}{\sqrt{\rho}}\right]-\frac{\hbar^{2}}{m^{2}}\int{d}^{4}x\sqrt{-g}\frac{\delta}{\delta{g_{\mu\nu}}}\Bigl(\lambda\frac{\nabla_{\sigma}\nabla^{\sigma}\sqrt{\rho}}{\sqrt{\rho}}\Bigr). \label{2ndcov0}
\end{eqnarray}
The first of these components cancel by virtue of Eq.~(\ref{CnstrEq}). Integration by parts can then be applied to the second term 
\begin{eqnarray}
\delta{A}_{cnstr}=\frac{\hbar^{2}}{m^{2}}\int{d}^{4}x\sqrt{-g}\frac{\delta}{\delta{g^{\mu\nu}}}\Biggl[\nabla_{\sigma}\sqrt{\rho}\nabla^{\sigma}\Bigl(\frac{\lambda}{\sqrt{\rho}}\Bigr)-\nabla_{\sigma}\Bigl(\lambda\frac{\nabla^{\sigma}\sqrt{\rho}}{\sqrt{\rho}}\Bigr)\Biggr]. \label{2ndcov1}
\end{eqnarray}
Here, the first and second components on the right-hand side will hereon be referred to as $\delta{A}_{cnstr,1}$ and $\delta{A}_{cnstr,2}$, respectively (i.e $\delta{A}_{cnstr}=\delta{A}_{cnstr,1}-\delta{A}_{cnstr,2}$). The Compton wavelength squared $\ell_{c}^{2}$ will be appended from these expressions for the time being. In applying the Laplace-Beltrami operator and expanding $\delta{A}_{cnstr,2}$, we have 
\begin{eqnarray}
\delta{A}_{cnstr,2}&=&\int{d}^{4}x\sqrt{-g}\frac{\delta}{\delta{g^{\mu\nu}}}\nabla_{\sigma}\Bigl(\lambda\frac{\nabla^{\sigma}\sqrt{\rho}}{\sqrt{\rho}}\Bigr)=\int{d}^{4}x\sqrt{-g}\frac{\delta}{\delta{g^{\mu\nu}}}\Bigl[\frac{1}{\sqrt{-g}}\partial_{\sigma}\Bigl(\sqrt{-g}\lambda\frac{\partial^{\sigma}\sqrt{\rho}}{\sqrt{\rho}}\Bigr)\Bigr] \label{2ndcov2} \\
&=&\int{d}^{4}x\sqrt{-g}\frac{\delta}{\delta{g^{\mu\nu}}}\Bigl(\frac{1}{\sqrt{-g}}\Bigr)\partial_{\sigma}\Bigl(\sqrt{-g}\lambda\frac{\partial^{\sigma}\sqrt{\rho}}{\sqrt{\rho}}\Bigr)+\int{d}^{4}x\sqrt{-g}\frac{1}{\sqrt{-g}}\frac{\delta}{\delta{g^{\mu\nu}}}\partial_{\sigma}\Bigl(\sqrt{-g}\lambda\frac{\partial^{\sigma}\sqrt{\rho}}{\sqrt{\rho}}\Bigr). \nonumber
\end{eqnarray}
Using the Jacobi identity, one can identify that $\frac{\delta}{\delta g^{\mu\nu}}{\sqrt{-g}}=-\frac{1}{2}g_{\mu\nu}{\sqrt{-g}}$. The inverse expression would then take the form $\frac{\delta}{\delta g^{\mu\nu}}\Bigr(\frac{1}{\sqrt{-g}}\Bigl)=\frac{1}{2}g_{\mu\nu}\frac{1}{\sqrt{-g}}$. Equation~(\ref{2ndcov2}) can then be written as
\begin{eqnarray}
\delta{A}_{cnstr,2}&=&\frac{1}{2}\int{d}^{4}x\sqrt{-g}g_{\mu\nu}\frac{1}{\sqrt{-g}}\partial_{\sigma}\Bigl(\sqrt{-g}\lambda\frac{\partial^{\sigma}\sqrt{\rho}}{\sqrt{\rho}}\Bigr)+\frac{\delta}{\delta{g^{\mu\nu}}}\int{d}^{4}x\partial_{\sigma}\Bigl(\sqrt{-g}\lambda\frac{\partial^{\sigma}\sqrt{\rho}}{\sqrt{\rho}}\Bigr) \nonumber \\
&=&\frac{1}{2}\int{d}^{4}x\Biggl[g_{\mu\nu}\partial_{\sigma}\Bigl(\sqrt{-g}\lambda\frac{\partial^{\sigma}\sqrt{\rho}}{\sqrt{\rho}}\Bigr)\Biggr]=\frac{1}{2}\int{d}^{4}x\sqrt{-g}\Biggl[g_{\mu\nu}\nabla_{\sigma}\Bigl(\lambda\frac{\nabla^{\sigma}\sqrt{\rho}}{\sqrt{\rho}}\Bigr)\Biggr]. \label{2ndcov3}
\end{eqnarray}
The second component on the right-hand side of the first line is a boundary term and disappears for infinitesimally small $\lambda$ and $\sqrt{\rho}$ at the boundary. As was demonstrated for a local Gaussian density, this would not happen within the framework of the linear-order constraint, rather only for the mediating field (Eq.~(\ref{eternal_eq})) derived from the exponential constraint. The other component within $\delta{A}_{cnstr}$ can also be trivially evaluated by accounting for the identity $\frac{\delta}{\delta{g^{\mu\nu}}}g^{\alpha\beta}=\frac{1}{2}(\delta_{\mu}^{\alpha}\delta_{\nu}^{\beta}+\delta_{\mu}^{\beta}\delta_{\nu}^{\alpha})$, so as to arrive at
\begin{eqnarray}
\delta{A}_{cnstr,1}=\int{d}^{4}x\sqrt{-g}\frac{\delta}{\delta{g^{\mu\nu}}}\nabla_{\sigma}\sqrt{\rho}\nabla^{\sigma}\Bigl(\frac{\lambda}{\sqrt{\rho}}\Bigr)=\frac{1}{2}\int{d}^{4}x\sqrt{-g}\Biggl[\nabla_{\mu}\sqrt{\rho}\nabla_{\nu}\Bigl(\frac{\lambda}{\sqrt{\rho}}\Bigr)+\nabla_{\nu}\sqrt{\rho}\nabla_{\mu}\Bigl(\frac{\lambda}{\sqrt{\rho}}\Bigr)\Biggr]. \label{2ndcov4}
\end{eqnarray}
In then adopting the brackets for the symmetrized permutation of indices in the above expression, we arrive at the acquired stress-energy tensor for the constraint
\begin{eqnarray}
\delta{A}_{cnstr}=\frac{\hbar^2}{2m^2}\int{d}^{4}x\sqrt{-g}\Biggl[\nabla_{(\mu}\sqrt{\rho}\nabla_{\nu)}\Bigl(\frac{\lambda}{\sqrt{\rho}}\Bigr)-g_{\mu\nu}\nabla_{\sigma}\Bigl(\lambda\frac{\nabla^{\sigma}{\sqrt{\rho}}}{\sqrt{\rho}}\Bigr)\Biggr]. \label{2ndcov5}
\end{eqnarray}
Q.E.D. 
\end{widetext}


\end{document}